\documentclass[aps,twocolumn]{revtex4}

\usepackage{setspace}
\usepackage{graphicx}

\begin{document}

\title{Geometric dephasing-limited Hanle effect in long-distance lateral
silicon spin transport devices}

\author{Biqin Huang}
\altaffiliation{bqhuang@udel.edu}
\author{Hyuk-Jae Jang}
\author{Ian Appelbaum}
\altaffiliation{Present address: Physics Department, University of Maryland (appeli@umd.edu)}
\affiliation{ Electrical and Computer Engineering Department,
University of Delaware, Newark, Delaware, 19716}

\begin{abstract}
Evidence of spin precession and dephasing (``Hanle effect'') induced by an external magnetic field is the only unequivocal proof of spin-polarized conduction electron transport in semiconductor devices. However, when spin dephasing is very strong, Hanle effect in a uniaxial magnetic field can be impossible to measure. Using a Silicon device with lateral injector-detector separation over 2 millimeters, and geometrically-induced dephasing making spin transport completely incoherent, we show experimentally and theoretically that Hanle effect can still be measured using a two-axis magnetic field. 

\end{abstract}

\maketitle
\newpage

The ``Hanle effect'', which describes the phenomenon of suppressed spin precession signal due to precession angle uncertainty (spin ``dephasing''), is the sole unambiguous method to confirm the presence of spin transport in solid state devices\cite{JOHNSON85, JOHNSON88, MONZON, LOU, JONKERLATERAL, FABIAN, DEPHASINGPRB}. In vertical-transport spintronic devices, dephasing is induced predominantly by diffusion and is relatively weak, resulting in clear spin precession signals that allow the measurement of long conduction electron spin lifetimes in silicon\cite{APPELBAUM, BIQIN350,BIQINJAP}. However, when dephasing becomes extremely strong (due to transit-time uncertainty induced by either diffusion or geometry), coherent spin precession can be fully suppressed and Hanle effect is difficult if not impossible to measure. 

To observe spin transport over a long distance, it is necessary to have a device with lateral transport geometry. In that case, geometrically-induced dephasing can be dominant because of a large lateral extent of injector and detector, causing a large transport-length uncertainty. In this report, we present a theoretical and experimental study of the Hanle effect in a system under conditions of extreme dephasing over 2 millimeters of lateral spin transport in undoped silicon. Our results show that spin transport can be confirmed with Hanle effect using two-axis magnetic field measurements even when the spins precess completely incoherently.

\begin{figure}
  \centering
  \includegraphics[scale=0.5]{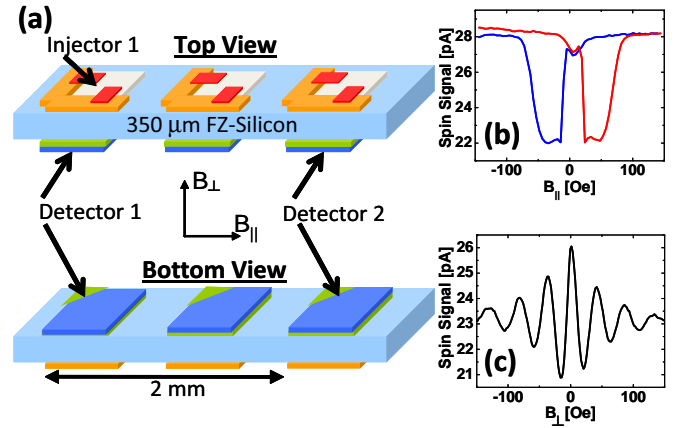}
  \caption{ \label{fig:fig1}
(a)Schematic illustration of lateral silicon spintronic device with spin injector on one side (``top view'') of a 350-micron-thick undoped Si (100) wafer, and spin detector on the other (``bottom view''). (b) In-plane spin valve effect with vertical transport from spin injector 1 to spin detector 1 at 85K. (c) Hanle effect measurement in a perpendicular magnetic field with at least 6 $\pi$ rad. oscillations indicates coherent transport vertically through 350 $\mu m$ distance, as in Ref. \onlinecite{BIQIN350}.
}
\end{figure}

Under the influence of an external magnetic field $B_\perp$ applied perpendicularly to the initial spin direction at the injector, spin precession is induced at an angular frequency $\omega=g\mu_B B_\perp/\hbar$ (where $g$ is the electron spin g-factor, $\mu_B$ is the Bohr magneton, and $\hbar$ is the reduced Planck constant). If transport occurs over a transit time $\tau$, the output at the detector will be dependent on the projection of the final spin direction onto the magnetization direction of the detector. The sum of the resulting $\cos{\omega\tau}$ contributions of electrons with different transit times (driven by diffusion or different transit lengths) can therefore cause at least partial signal cancellation. This Hanle effect is particularly difficult to measure in the presence of complete dephasing (caused by a very wide transit-time distribution), because the signal is completely suppressed except when $B_\perp=0$.

\begin{figure}
  \centering
  \includegraphics[scale=0.7]{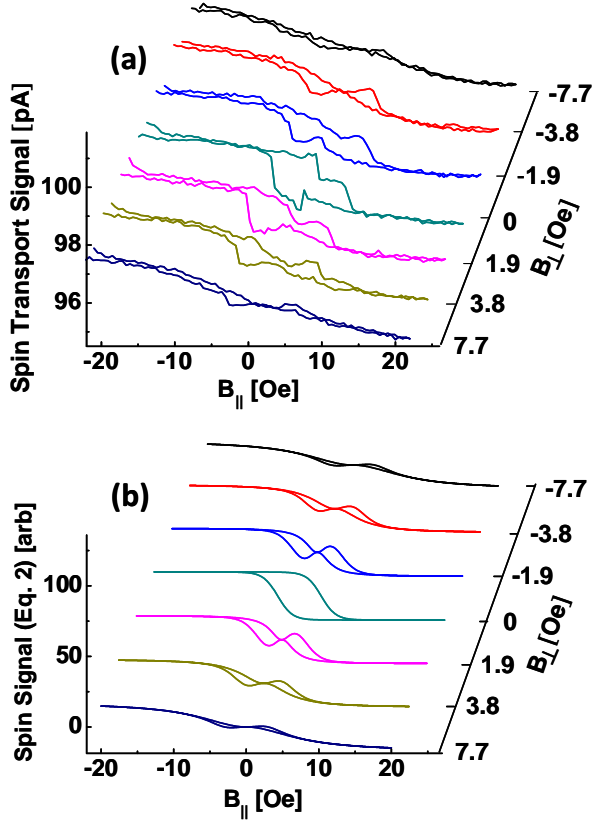}
  \caption{ \label{fig:fig3}
(a) In-plane minor-loop measurements in a static perpendicular magnetic field $B_\perp$. (b) Simulated minor loops using the incoherent spin transport model, Eq. (\ref{DEPHASE}). The increase of perpendicular magnetic field magnitude leads to a collapse of the minor loop due to precession and strong dephasing.}
\end{figure}

\begin{figure}
  \centering
  \includegraphics[scale=0.6]{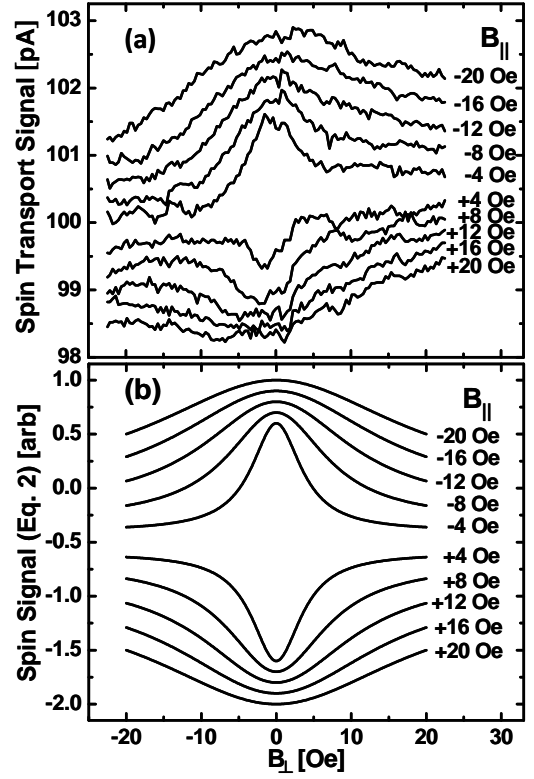}
  \caption{ \label{fig:fig4}
(a) Perpendicular-field Hanle effect measurements in a static in-plane magnetic field $B_{||}$. The slight asymmetry in peak width for positive and negative $B_{||}$ (which causes magnetization switching and hence flips the signal) indicates a small $B_{||}$ uncertainty.(b) Simulations using Eq. (\ref{DEPHASE}) show excellent agreement with the experiments. Curves have been offset for clarity.}
\end{figure}

However, when a static in-plane field $B_{||}$ is applied along with $B_\perp$, the output spin signal due to each electron will be proportional to\cite{OBLIQUE}:

\begin{equation}\label{MHANLE}
\frac{B_\perp^2\cos{\omega \tau}+B_{||}^2}{B_\perp^2+B_{||}^2},
\end{equation}

\noindent where in this case $\omega=g\mu_B \sqrt{B_\perp^2+B_{||}^2}/\hbar$.

Because of the two-axis field configuration, the contributions to the precession signal have now been decomposed into two terms: the first, dependent on spin precession at frequency $\omega$, is the coherent part, and the second term (determined solely by field geometry) is the incoherent part. For completely incoherent transport induced by dominant dephasing, the first term is averaged out by the wide transit-time distribution and the expected output signal is simply proportional to 

\begin{equation}\label{DEPHASE}
\frac{B_{||}^2}{B_\perp^2+B_{||}^2}.
\end{equation}

\noindent This expression is simply the projection of a homogeneous cone of spins having an axis along $\vec{B_\perp}+\vec{B_{||}}$. 

Of particular importance is when $B_{||}=0$. Then, this expression predicts that the signal is zero, except when $B_\perp$ is also identically zero, making standard Hanle impossible to measure. It is important to note here that we have assumed that the spin lifetime is much longer than the transport time, which is the case in silicon devices at low temperature where the lifetime can easily exceed several hundred nanoseconds\cite{BIQIN350}.

To construct a silicon device with a long lateral spin transport distance and extreme geometrically-induced dephasing, fabrication techniques similar to the ones adopted in vertical spintronic devices were used\cite{APPELBAUM, BIQIN350,BIQINJAP}. Under ultra-high vacuum, a float-zone-grown undoped Si(100) wafer (FZ-Si) used for spin transport was bonded with a metal thin film to a Silicon-on-Insulator (SOI) wafer with n-type 3$\mu m$ device layer(1-20 $\Omega \cdot cm$) $\it{in} \it{situ}$, resulting in a bonded structure of FZ-Si/Ni$_{80}$Fe$_{20}$(4 nm)/Cu(4 nm)/SOI. The subsequently exposed 3$\mu m$-thick device layer was then patterned into hot-electron detectors as shown in Fig. 1(a), and isolated with shallow trenches (a few microns deep into FZ-Si). The resulting spin detectors therefore have an inverted geometry in comparison to our vertical-transport devices. Finally, 40 nm Al/10 nm Co$_{84}$Fe$_{16}$/Al$_2$O$_3$/7 nm Al /10nm Cu spin-injector tunnel junctions were fabricated on the opposite side of the wafer.\cite{35PERCENT} Fig. 1(a) shows schematic top and bottom views clarifying the device structure. The tunnel junction injects spin-polarized ballistic hot electrons over the Cu/FZ-Si Schottky barrier, and spin-dependent scattering in the Ni$_{80}$Fe$_{20}$ bonding layer is used to detect the final spin polarization orientation after transport by measuring the ballistic current passing into the conduction band of the bonded n-Si SOI device layer detector. Further details on these hot-electron techniques can be found in Refs. \onlinecite{APPELBAUM}, \onlinecite{BIQIN350}, and \onlinecite{35PERCENT}.

In comparison to previously-demonstrated vertical-transport devices, this device architecture enables the electron spin to travel in a lateral direction by utilizing different injector and detector combinations. First, to determine the basic device performance, a vertical configuration was used by injecting spin polarized electrons from injector 1 and detecting the electron spin at detector 1 directly below it, using an injector tunnel-junction voltage of -1.0 V and an injector-detector accelerating voltage of 30V at 85K. Spin-valve measurements in an in-plane magnetic field, shown in Fig. 1(b), demonstrate a high magneto-current ratio (about 28\%, corresponding to approximately 12\% current spin polarization) and indicates the high quality of the spin injector. Since there is no geometrically-induced dephasing in this configuration (transport distance is the same for all electrons), Fig. 1(c) shows multiple precession oscillations in a perpendicular magnetic field, and a suppression of signal at higher fields (the Hanle effect due to dephasing), which proves the existence of coherent vertical spin transport. 

In contrast, by using detector 2 (displaced by 2 millimeters from detector 1), the transport is now primarily lateral. Because the injector and detector have lateral extent of 500 and 1000 microns respectively, this configuration now results in a large transit-length uncertainty, which induces strong dephasing. Under these conditions, standard spin precession/Hanle effect measurement with a purely perpendicular single-axis measurement is precluded because the signal is canceled by a fully dephased ensemble of arriving electrons. However, examination of Eq. (\ref{DEPHASE}) shows us that an in-plane field $B_{||}$ can be used to uncover the presence of spin transport {\it{even if it is entirely incoherent}. }

A series of in-plane magnetic field minor-loop measurements (varying $B_{||}$ so that only the magnetically softer NiFe layer in the detector switches magnetization) were done with a tunnel-junction voltage of -1.2V and injector-to-detector voltage of 160V at 50K in different static perpendicular magnetic field -7.7Oe$<B_\perp<$+7.7Oe. As can be seen in Fig. 2(a), except when $B_\perp=0$, the spin transport signal is suppressed for $B_{||}\approx 0$ where the hysteresis loop collapses because of spin dephasing. Furthermore, the characteristic width of this suppressed region is larger for larger values of $B_\perp$, independent of its polarity. This can be understood simply by considering that out-of-plane fields induce strong signal cancellation from dephased precessing spins, and it takes larger values of in-plane fields to tilt the total magnetic field $\vec{B}_\perp+\vec{B}_{||}$ toward the in-plane measurement axis as $B_\perp$ increases.

The prediction of the model given in Eq. (\ref{DEPHASE}) for the same field configuration (and incorporating the detector NiFe magnetization switching) is shown for comparison in Fig. 2(b). The qualitative and quantitative agreement unambiguously confirm spin transport over 2 millimeters of Silicon, despite the fully incoherent spin precession caused by geometry of the injector and detector.

Similarly, Hanle effect measurements at different static in-plane magnetic fields can be done. The results for $B_{||}$ at nominal fields from -20 Oe to 20 Oe are shown in Fig. 3(a). With a decrease of the in-plane magnetic field, the width of the Hanle peak decreases because it becomes easier to orient the total magnetic field out-of-plane where dephasing is strongest. These experimental results agree well with the prediction of Eq. (\ref{DEPHASE}), as shown in Fig. 3(b).

In summary, a two-axis magnetic field configuration can be used to confirm spin transport with evidence for spin precession in a semiconductor device, despite the presence of extreme spin dephasing which would otherwise have made single-axis measurements impossible. We have demonstrated this technique by comparing a simple model to results from a 2 millimeter transport distance device fabricated from silicon. Because there is therefore always a means of measuring the effects of spin precession when {\it bona fide} spin transport is present, we conclude that there is no reason to rely on spin-valve measurements in spintronics devices, where spurious effects can obscure the true transport mechanism.\cite{MONZON}

\end{document}